\documentclass[preprint,aps,prd,showpacs,nofootinbib,superscriptaddress,12pt]{revtex4-1}
\usepackage{amssymb}

\usepackage{graphicx}
\graphicspath{{figures/}{fig/}}
\usepackage{amsmath}
\usepackage{bm}
\usepackage{slashed}
\usepackage{epsfig}
\usepackage{amsfonts}
\usepackage{epstopdf}
\usepackage{color}
\usepackage{extarrows}
\usepackage{multirow}
\usepackage[utf8]{inputenc}
\usepackage{hyperref}

\newcommand{\ben}{\begin{eqnarray}}
\newcommand{\een}{\end{eqnarray}}

\newcommand{\bef}{\begin{figure}[!htp]}
\newcommand{\eef}{\end{figure}}

\newcommand{\bea}{\begin{eqnarray}}
\newcommand{\eea}{\end{eqnarray}}

\def\ba{\begin{linenomath*}\begin{equation}}
\def\ea{\end{equation}\end{linenomath*}}

\allowdisplaybreaks

\newcommand{\state}[4]{{^{#1}\hspace{-0.6mm}#2_{#3}^{[#4]}}}
\newcommand{\stateprime}[4]{{^{#1}\hspace{-0.6mm}#2_{#3}^{\prime[#4]}}}

\newcommand\CScPz{\state{3}{P}{0}{1}}

\newcommand\COaSz{\state{1}{S}{0}{8}}

\newcommand\lrd{\overleftrightarrow{D}}

\begin{document}

\title{The single parton fragmentation functions of heavy
quarkonium in soft gluon factorization}
\author{Qi-Lin Jia}
\email{qilinjia@jxnu.edu.cn}
\author{An-Ping Chen}
\email{chenanping@jxnu.edu.cn}
\author{Yuan-Guo Xu}
\email{yuanguoxu@jxnu.edu.cn}

\affiliation{ College of Physics and Communication Electronics, Jiangxi Normal University, Nanchang 330022, China}

\date{\today}

\begin{abstract}
We study the single parton fragmentation functions (FFs) at the input factorization scale $\mu_0\gtrsim 2m_Q$, with heavy quark mass $m_Q$, in the soft gluon factorization (SGF) approach. We express the FFs in terms of perturbatively calculable short distance hard parts for producing a heavy quark-antiquark pair in all possible states, convoluted with corresponding soft gluon distribution for the hadronization of the pair to a heavy quarkonium. We compute the perturbative
short distance hard parts for producing a heavy quark pair in all possible $S$-wave and $P$-wave states up to $O(\alpha_s^2)$. With
our results, the SGF can be further used to study the heavy quarkonium production at the hadron colliders and heavy quarkonium
production within a jet.
\end{abstract}

\maketitle
\allowdisplaybreaks

\section{INTRODUCTION}

As the simplest bound state of strong interactions, heavy quarkonium provides an ideal physical system to explore both perturbative and nonperturbative aspects of quantum chromodynamics (QCD). The successful performance of the Tevatron and the LHC has further heightened interest in studying heavy quarkonium production at hadron colliders.

In the high transverse momentum ($p_T$) region, the cross section for heavy quarkonium hadroproduction can be factorized using the collinear factorization formalism~\cite{Kang:2014tta,Kang:2014pya},
\begin{align}\label{eq:pqcdfac}
\mathrm{d}\sigma_{A+B\to H+X}(p) \approx &
\sum_{i,j}f_{i/A}(x_1,\mu_F)f_{j/B}(x_2,\mu_F)
 \Big \{ \sum_{f} D_{f\to H}(z,\mu_F)
    \otimes
\mathrm{d}{\hat{\sigma}}_{i+j\to f+X}({\hat P}/z,\mu_F)
 \nonumber\\
&\hspace{-3cm} +  \sum_{\kappa}  {\cal
D}_{[Q\bar{Q}(\kappa)]\to H}(z,\zeta,\zeta',\mu_F)
 \otimes
\mathrm{d}{\hat{\sigma}}_{i+j\to [Q\bar{Q}(\kappa)]+X}({\hat P}(1\pm\zeta)/2z,{\hat
P}(1\pm\zeta')/2z,\mu_F)   \Big \},
\end{align}
where  $\mathrm{d}{\hat{\sigma}}$'s are perturbative calculable hard parts describing partonic interactions. $f_{i/A}$ denotes the  parton distribution function (PDF), while  $D_{f\to H}$ is the single parton fragmentation function (FF) which gives the leading power (LP) contribution in $1/p_T$ expansion, ${\cal D}_{[Q\bar{Q}(\kappa)]\to H}$ is the double parton FF~\cite{Kang:2014tta,Kang:2014pya} that gives the next-to-leading power (NLP) contribution.$\sum_f$ runs over all parton flavors, and the  $\sum_\kappa$ runs over all possible spin and color states of the fragmenting $Q\bar{Q}$-pair.
In this context, $p$ denotes the momentum of the observed heavy quarkonium,  $\hat{P}^\mu = (p^+,0,\vec{0}_\perp)$ is a light like momentum whose plus component equals to the plus component of $p^\mu$,  $z$, $\zeta$ and $\zeta'$  represent the light-cone momentum fractions, and $\mu_F$ denotes the collinear factorization scale. The dependence of the fragmentation functions on $\mu_F$ follows the evolution equations \cite{Gribov:1972ri,Altarelli:1977zs,Dokshitzer:1977sg,Kang:2014tta}. With input FFs at a initial scale $\mu_0\gtrsim 2m_Q$, where $m_Q$ represents the mass of the heavy quark, the FFs at other scales
can be obtained through the evolution.
The input FFs are nonperturbative and, in principle should be determined from experiments.
However, due to $\mu_0\gg \Lambda_{\textrm{QCD}}$, it is reasonable to further factorize these input FFs  using  non-relativistic QCD (NRQCD) factorization \cite{Bodwin:1994jh} and the soft gluon factorization (SGF) \cite{Ma:2017xno}.

NRQCD factorization is both the most theoretically sound and phenomenologically successful theory in describe the quarkonium production so far. In NRQCD factorization, the FFs are factorized into summation of perturbatively calculable short-distance coefficients (SDCs) multiplied by nonperturbative long-distance matrix elements (LDMEs). The NRQCD factorization for FFs has been extensively studied. The SDCs for all double parton FFs have been calculated up to $O(\alpha_s)$ in Refs. \cite{Ma:2013yla,Ma:2014eja,Ma:2015yka}.
The SDCs for all single parton FFs are available up to $O(\alpha_s^2)$\cite{Beneke:1995yb,Braaten:1993mp,Braaten:1993rw,Cho:1994gb,Braaten:1994kd,Ma:1995vi,Braaten:1996rp,Braaten:2000pc,Hao:2009fa,Jia:2012qx,Bodwin:2014bia}
(see \cite{Ma:2013yla,Ma:2014eja,Ma:2015yka} for a summary and comparison). And part of them are calculated to $O(\alpha_s^3)$ \cite{Zhang:2017xoj,Braaten:1993rw,Braaten:1995cj,Bodwin:2003wh,Bodwin:2012xc, Zhang:2018mlo,Artoisenet:2018dbs,Feng:2018ulg, Zhang:2020atv,Zheng:2021ylc,Feng:2021uct}.

However, recent studies shown that NRQCD factorization encounters some difficulties
in describing inclusive quarkonium production data.  \cite{Campbell:2007ws, Artoisenet:2007xi,Ma:2008gq,Gong:2009kp, Zhang:2009ym,
Gong:2008sn, Gong:2008hk,Li:2011yc, Butenschoen:2012px, Chao:2012iv,
Gong:2012ug, Butenschoen:2014dra, Han:2014jya, Zhang:2014ybe,Feng:2018ukp,Chen:2022qli,Chen:2023gsu}. To overcome these difficulties, the SGF approach has been proposed. It was argued that the SGF is
equivalent to the NRQCD factorization, but with a series of important relativistic corrections originated from kinematic effects resummed \cite{Chen:2020yeg}. In Refs. \cite{Chen:2022qli,Chen:2023gsu}, SGF has been applied to study color-octet contributions in the $J/\psi$ inclusive production at B factories and the $\chi_{cJ}$ production at LHC with large $p_T$. These phenomenological studies demonstrate that SGF not only alleviates the universality problem but also resolves the issue of negative cross sections in NRQCD factorization.  To further apply SGF to heavy quarkonium hadroproduction, it is necessary to study the FFs using the SGF approach.

In SGF, the FFs are factorized as a form of perturbative short-distance hard part convoluted with soft gluon distribution (SGD). Part of the short-distance hard parts for single parton FFs have been calculated up to $O(\alpha_s^2)$ in Refs. \cite{Chen:2021hzo,Chen:2023gsu}. And part of the short-distance hard part of double parton FFs has been obtained at leading order (LO) in Ref. \cite{Chen:2023gsu}. In this paper we will derive complete contributions to the short-distance hard parts for single parton FFs at $O(\alpha_s^2)$.

The structure of this paper is as follows: In Sec. \ref{sec:SGF}, we introduce the SGF formula for single parton FFs, including the definition of SGDs for different states. In Sec. \ref{sec:HP}, we compute the related short-distance hard parts. We summarize our results in Sec. \ref{sec:summary}.

\section{FRAGMFNTATION FUNCTIONS IN SGF}\label{sec:SGF}

According to Refs. \cite{Ma:2017xno,Chen:2021hzo}, in SGF the single
parton FFs at scale $\mu_0$ can be factorized as
\begin{align}\label{eq:SGF-form}
D_{f \rightarrow H}(z,\mu_0)
=&  \sum_{n,n^\prime} \int \frac{\mathrm{d}x}{x}   \hat{D}_{ f \to Q\bar{Q}[nn^\prime] }(\hat{z}; M_H/x, m_Q,\mu_0, \mu_\Lambda)
F_{[nn^\prime] \to H}(x,M_H,m_Q,\mu_\Lambda),
\end{align}
where $\hat z=z/x$, $\mu_\Lambda$ is the factorization scale, $\hat{D}_{ f \to Q\bar{Q}[nn^\prime] }$ refers to the perturbatively calculable short-distance hard parts  that produce a $Q \bar Q$ pair with quantum numbers $n=\state{{2S+1}}{L}{J,J_z}{c}$ and $n^\prime =\stateprime{{2S^\prime+1}}{L}{J^\prime,J_z^\prime}{c^\prime}$ in the amplitude and the complex-conjugate of the amplitude, respectively. $M_H$ is the mass of heavy quarkonium $H$ which satisfies $p^2=M_H^2$.
$F_{[nn^\prime] \to H}$ is the SGD, which describes the hadronization of an intermediate $Q\bar Q$ pair into heavy quarkonium by radiate soft gluons. The SGDs are defined as:
\begin{align}\label{eq:SGD-1d}
F_{[nn^\prime] \to H}(x,M_H,m_Q,\mu_\Lambda)
&= p^+\int \frac{\mathrm{d}b^-}{2\pi} e^{-ip^+  b^-/x}  \langle 0| [\bar\Psi \mathcal {K}_{n} \Psi]^\dag(0) [a_H^\dag a_H] [\bar\Psi \mathcal {K}_{n^\prime}\Psi](b^-) |0\rangle_{\textrm{S}},
\end{align}
where $x$ is the light-cone momentum fraction which defined as $x = p^+/P_c^+$, and $P_c$ is the total momentum of the intermediate $Q\bar Q$ pair. $\Psi$ stands for Dirac field of heavy quark and the subscript ``S'' indicates that the field operators in the definition are obtained in the small momentum region. In additional, we define ``S'' to select only leading power terms in $(P_c-p)^+=(1-x)P_c^+$ expansion \cite{Chen:2021hzo}. $\mathcal {K}_{n}$ are projection operators corresponding to the intermediate state $n$. In this paper, we are interested in $n=\state{3}{S}{1,S_z}{1,8}$, $\state{1}{S}{0}{1,8}$, $\state{3}{P}{J,J_z}{1,8}$ and $\state{1}{P}{1,J_z}{1,8}$. For these states we have~\cite{Ma:2017xno}
\begin{align}
 \mathcal {K}_{n}(b^-) =& \frac{\sqrt{M_ {H}}}{M_ {H}+2 m_Q}\frac{ M_{H} + \slashed{p}}{2M_ {H}} \Gamma_n  \frac{M_{H} - \slashed{p}}{2M_{H}},
\end{align}
with
\begin{subequations}\label{eq:spinProj}
\begin{align}
	\Gamma_n  =&
	\epsilon_{S_z}^\mu \gamma_\mu \mathcal {C}^{[c]}, &\text{for $n= \state{3}{S}{1,S_z}{c}$,}  \\
	\Gamma_n  =&  \gamma_5 \mathcal {C}^{[c]}, &\text{for $n= \state{1}{S}{0}{c}$,}
 \\
	\Gamma_n  =&  \mathcal {E}_{J,J_z}^{\mu\nu} \gamma_\mu \Big( -\frac{i}{2}  \overleftrightarrow{D}_\nu \Big) \mathcal {C}^{[c]}, &\text{for $n= \state{3}{P}{J,J_z}{c}$,}
 \\
	\Gamma_n  =&  \gamma_5 \epsilon_{J_z}^\mu \Big( -\frac{i}{2}  \overleftrightarrow{D}_\mu \Big) \mathcal {C}^{[c]}, &\text{for $n= \state{1}{P}{1,J_z}{c}$.}
\end{align}
\end{subequations}
The color operators $\mathcal {C}^{[c]}$ in above are defined as
\begin{subequations}
\begin{align}
\mathcal {C}^{[1]} =&\frac{{\bm 1}_c}{\sqrt{N_c}},\\
\mathcal {C}^{[8]} =& \sqrt{2}T^{ \bar a} \Phi_{l}(rb^-)_{\bar a a},
\end{align}
\end{subequations}
where ${\bm 1}_c$ represents the identity matrix in the color space, and $T^{ \bar a}$ is the generator of the fundamental (triplet) representation of the $\textrm{SU}(3)$ gauge group. The gauge link $\Phi_{l}(rb^-)_{\bar a   a}$ is introduced to ensure gauge invariance of the SGDs. And the gauge link is defined along the $l^\mu = (0, 1, \vec{0}_\perp)$ direction,
\begin{equation}\label{eq:gaugelink}
  \Phi_l(rb^{-})= \mathcal {P} \, \text{exp} \left[-i g_s
  \int_{0}^{\infty}\mathrm{d}\xi l\cdot A(rb^{-} + \xi l) \right] \, ,
\end{equation}
where $\mathcal {P}$ denotes path ordering, $A^{\mu}(x)$ is the matrix-valued gluon field in the adjoint representation: $[A^{\mu}(x)]_{ac} = i f^{abc} A^{\mu}_{b}(x)$.
In Eq.~\eqref{eq:spinProj}, $D_\mu$ is the gauge covariant derivative with $\overline\Psi \lrd_\mu \Psi =
\overline\Psi (D_\mu \Psi) -
(D_\mu \overline\Psi)\Psi$. And $\epsilon_{S_z}$, $\mathcal {E}_{J,J_z}$, $\epsilon_{J_z}$ represents the polarization tensors for $\state{3}{S}{1,S_z}{8}$ state, $\state{{3}}{P}{J,J_z}{1,8}$ state and $\state{{1}}{P}{1,J_z}{1,8}$ state. Among them, $\mathcal {E}_{J,J_z}$ can be expressed as
\begin{align}\label{eq:EJJz}
\mathcal {E}_{J,J_z}^{\mu\nu}
 =& \sum_{L_z,S_z^\prime} \langle 1,L_z;1,S_z^\prime | J,J_z \rangle \epsilon_{S_z^\prime}^\mu\epsilon_{L_z}^\nu,
\end{align}
where $\langle 1,L_z;1,S_z^\prime | J,J_z \rangle$ is the Clebsch-Gordan coefficient, $\epsilon_{L_z}$ is the polarization vector for the $P$-wave orbital angular momentum state.
In Eq.~(\ref{eq:SGF-form}), it was suggested to expanding $m_Q^2$ around $M_H^2/(4x^2)$ in the short distance hard parts \cite{Ma:2017xno,Chen:2021hzo},
\begin{align}\label{eq:velocity-expansion}
\hat{D}_{ f \to Q\bar{Q}[nn^\prime] }(\hat{z}; M_H/x, m_Q,\mu_0, \mu_\Lambda)
 =& \sum_{i=0} \hat{D}_{ f \to Q\bar{Q}[nn^\prime] }^{(i)}(\hat{z}; M_H/x, \mu_0, \mu_\Lambda)
\biggr(m_Q^2-\frac{M_H^2}{4x^2}\biggr)^i,
\end{align}
which defines a velocity expansion in SGF. Here we only consider the hard parts at leading order in the velocity expansion, then we have $n=n^\prime$. For convenient, we denote
\begin{align}
[\state{{2S+1}}{L}{J,\lambda}{c}] \equiv [\state{{2S+1}}{L}{J,\lambda}{c}\state{{2S+1}}{L}{J,\lambda}{c}].
\end{align}

Following Ref. \cite{Ma:2015yka}, to study the $\state{{3}}{P}{J,J_z}{8}$ contributes to the production and polarization of quarkonium like $\psi(nS)$ and $\Upsilon(nS)$, it is convenient to define the $\state{3, S_z}{P}{}{1,8}$ channels, in which the spin of the intermediate $Q\bar Q$ pair is $S_z$, and the orbital angular momentum $L_z$ is summed over. The corresponding SGDs are defined by
\begin{align}
F_{[\state{3, S_z}{P}{}{c}] \to H}(x,M_H,m_Q,\mu_\Lambda)=& \sum_{J,J_z,S_z^\prime=S_z} F_{[\state{3}{P}{J,J_z}{c}] \to H}(x,M_H,m_Q,\mu_\Lambda)
\nonumber\\
 &\hspace{-4.5cm}=
 \frac{\mathbb{P} ^{\alpha \beta}p^+}{d-1}\int \frac{\mathrm{d}b^-}{2\pi} e^{-ip^+  b^-/x} \nonumber\\
&\hspace{-4.5cm}~~~~ \times \langle 0| [\bar\Psi \epsilon_{S_z}^\mu  \gamma_\mu ( -\frac{i}{2}  \overleftrightarrow{D}_\alpha ) \mathcal {C}^{[c]} \Psi]^\dag(0) [a_H^\dag a_H] [\bar\Psi  \epsilon_{S_z}^\nu \gamma_\nu ( -\frac{i}{2}  \overleftrightarrow{D}_\beta ) \mathcal {C}^{[c]} \Psi](b^-) |0\rangle_{\textrm{S}},
\end{align}
with
\begin{align}
\mathbb{P} ^{\alpha \beta}& =\sum_{L_z} \epsilon ^{\beta}_{L_z} \epsilon ^{*\alpha} _{L_z}= -g^{\alpha \beta}+\frac{p^\alpha p^{\beta}}{p^2}.
\end{align}

On the other hand, according to the discussion in Ref.~\cite{Ma:2015yka}, for the polarization of qurakonium production at hadron colliders, the polar angular distribution of the decaying products from the produced heavy quarkonium $H$ will depend only on the combination of the SGDs:
\begin{align}
\frac{1}{2}\Big(F_{[^{2S+1}L_{J,J_z}^{[c]}] \to H}(x,M_H,m_Q, \mu_\Lambda)+F_{[^{2S+1}L_{J,-J_z}^{[c]}] \to H}(x,M_H,m_Q, \mu_\Lambda)\Big).\nonumber
\end{align}
Therefore, similar to the polarized NRQCD LDMEs defined in Ref.~\cite{Ma:2015yka}, it is convenient to define polarized SGDs as follows:
\begin{align}
 F_{[n_\lambda] \to H}(x,M_H,m_Q, \mu_\Lambda)  =& \frac{1}{N_{n_\lambda}}\sum_{\vert J_z \vert = \lambda } F_{[^{2S+1}L_{J,J_z}^{[c]}] \to H}(x,M_H,m_Q, \mu_\Lambda),
\end{align}
where $n_\lambda$ denotes $^{2S+1}L_{J,\lambda }^{[c]}$, $\lambda=L, T, TT, \cdots$ correspond to $\vert J_z \vert=0, 1, 2, \cdots$, respectively. $\lambda=L$ represents
longitudinally polarized and $\lambda=T$ represents transversely polarized.  $N_{n_\lambda}$ is the number of polarization states for $n_\lambda$. We have~\cite{Ma:2015yka}
\begin{align}\label{eq:N-pol}
N_{^3{S}_{1,L}^{[c]}} &=N_{^1S_0^{[c]}}=N_{^1P_{1,L}^{[c]}}=N_{^3P_{0}^{[c]}}= N_{^3P_{1,L}^{[c]}}= N_{^3P_{2,L}^{[c]}}=N_{\state{3, L}{P}{}{c}}=1,\nonumber\\
N_{^3{S}_{1,T}^{[c]}} &=N_{^1P_{1,T}^{[c]}}= N_{^3P_{1,T}^{[c]}}= N_{^3P_{2,T}^{[c]}}=N_{\state{3, T}{P}{}{c}}=d-2,\nonumber\\
N_{^3P_{2,TT}^{[c]}}&=\frac{1}{2}(d-1)(d-2)-1,
\end{align}
where $d$ is the space-time dimension.

\section{The Short Distance Hard Parts}\label{sec:HP}

Following the matching procedure, to determine the
short distance hard part in Eq. \eqref{eq:SGF-form}, we replace the quarkonium $H$ by a on-shell $Q\bar{Q}$ pair with certain quantum number $n$ and momenta
\begin{align}
p_Q= \frac{1}{2}p +q ,  \quad \quad p_{\bar Q}= \frac{1}{2}p -q,
\end{align}
where $q$ is half of the relative momentum of the $Q\bar{Q}$ pair.
On-shell conditions $p_Q^2= p_{\bar Q}^2=m_Q^2$ result in
\begin{align}
p \cdot q= 0 , \quad \quad q^2=m_Q^2-p^2/4.
\end{align}
To project the final-state $Q \bar Q$ pair to the state $n$, we replace spinors of the $Q \bar Q$ by the projector~\cite{Ma:2017xno}
\begin{align}\label{eq:projector}
 \int \frac{\mathrm{d}^{d-2}\Omega}{N_\Omega} \frac{2}{\sqrt{M_H}(M_H+2m_Q)} ( \slashed{p}_{\bar Q} - m_Q )   \frac{M_H - \slashed{p}}{2M_H}  \tilde{\Gamma}_n
\frac{M_H + \slashed{p}}{2M_H} (\slashed{p}_{Q} +m_Q ),
\end{align}
here $\Omega$ denotes the solid angle of relative momentum $\textbf{q}$ in the $Q \bar Q$ rest frame, and $N_\Omega$ is given by
\begin{align}
N_\Omega = \int \mathrm{d}^{d-2}\Omega.
\end{align}
For different states $n$, the operators $\tilde{\Gamma}_n$ are given by
\begin{subequations}
\begin{align}
	\tilde{\Gamma}_n  =&
	\epsilon_{S_z}^{\ast \mu} \gamma_\mu \tilde{\mathcal {C}}^{[c]}, &\text{for $n= \state{3}{S}{1,S_z}{c}$,}  \\
	\tilde{\Gamma}_n  =&  \gamma_5 \tilde{\mathcal {C}}^{[c]}, &\text{for $n= \state{1}{S}{0}{c}$,}
 \\
	\tilde{\Gamma}_n  =&  \frac{(d-1) q_\alpha }{ \textbf{q}^2}  \mathcal {E}_{J,J_z}^{\ast \alpha \mu}\gamma_\mu \tilde{\mathcal {C}}^{[c]}, &\text{for $n= \state{3}{P}{J,J_z}{c}$,} \\
	\tilde{\Gamma}_n  =&  \frac{(d-1) q_\alpha }{ \textbf{q}^2} \epsilon_{J_z}^{\ast \alpha} \gamma_5 \tilde{\mathcal {C}}^{[c]}, &\text{for $n= \state{1}{P}{1,J_z}{c}$,}
\end{align}
\end{subequations}
where $\textbf{q}^2=-q^2$, and
\begin{subequations}
\begin{align}
\tilde{\mathcal {C}}^{[1]} =&\frac{{\bm 1}_c}{\sqrt{N_c}},\\
\tilde{\mathcal {C}}^{[8]} =& \sqrt{\frac{2}{N_c^2-1}}T^{  a} .
\end{align}
\end{subequations}

Here we use superscripts ``$LO$" and ``$NLO$" to denote the contributions at $O(\alpha_s)$ and $O(\alpha_s^2)$. Then inserting the perturbative expansions
\begin{subequations}
\begin{align}
D_{f \to Q\bar Q[n]} & = D^{LO}_{f \to Q\bar Q[n]} + D^{NLO}_{f \to Q\bar Q[n]} + \cdots, \\
\hat D_{f\to Q\bar Q[n^\prime]} &= \hat D^{LO}_{f\to Q\bar Q[n^\prime]}+ \hat D^{NLO}_{f\to Q\bar Q[n^\prime]} + \cdots, \\
F_{[n^\prime]\to Q\bar Q[n]} &= F^{LO}_{[n^\prime]\to Q\bar Q[n]} + F^{NLO}_{[n^\prime]\to Q\bar Q[n]} +\cdots,
\end{align}
\end{subequations}
into Eq. \eqref{eq:SGF-form} and using the orthogonal rations \cite{Ma:2017xno}
\begin{align}
F^{(0)}_{[n^\prime] \to Q\bar Q[n]}(x,M_H,m_Q,\mu _\Lambda )=\delta_{n' n} \delta(1-x),
\end{align}
we can obtain the matching relations for
the short distance hard parts up to $O(\alpha_s^2)$,
\begin{subequations}\label{eq:matching}
\begin{align}
\hat{D}_{f\to Q\bar Q[n]}^{LO,(0)}(z;M_H,\mu_0, \mu _\Lambda )& = D^{LO}_{f\to Q\bar Q[n]}(z; M_H, m_Q, \mu_0)\Big \vert_{m_Q^2 = M_H^2/4},\\
\hat{D}_{f\to Q\bar Q[n]}^{NLO,(0)}(z;M_H,\mu_0, \mu _\Lambda )&= \Big[ D^{NLO}_{f\to Q\bar Q[n]}(z;M_H,m_Q,\mu _\Lambda ) \nonumber\\
&\hspace{-3.5cm} -\sum_{n^\prime} \int \frac{\mathrm{d} x}{x} \hat{D}_{f\to Q\bar Q[n^\prime]}^{LO}(\hat z;M_H/x,m_Q,\mu_0,\mu _\Lambda ) F^{NLO}_{[n^\prime]\to Q\bar Q[n]}(x, M_H,m_Q,\mu_\Lambda ) \Big] \Big \vert_{m_Q^2 = M_H^2/4}.
\end{align}
\end{subequations}
Using these relations, we can match the
perturbative calculated SGDs to FFs to obtain the short distance hard parts.

In our calculation, we utilize the following projection operators to sum over the polarizations of different states \cite{Ma:2015yka},
\begin{subequations}\label{eq:projection operator}
\begin{align}
&\mathbb{P}_{0}^{\beta\beta^\prime \sigma\sigma^\prime} \equiv
 \sum_{|J_z|=0}\mathcal {E}_{0,J_z}^{\beta\sigma} \mathcal {E}_{0,J_z}^{\ast \beta ^\prime\sigma^\prime}   = \frac{1}{d-1} \mathbb{P}^{\beta\sigma}\mathbb{P}^{\beta^\prime\sigma^\prime},\\
& \mathbb{P}_{1,T}^{\beta\beta^\prime \sigma\sigma^\prime} \equiv \sum_{|J_z|=1}\mathcal {E}_{1,J_z}^{\beta\sigma} \mathcal {E}_{1,J_z}^{\ast \beta ^\prime\sigma^\prime}   \nonumber\\&  =\frac{1}{2}
\Big(  \mathbb{P}_{\perp}^{\beta\beta^\prime}\mathbb{P}_{\parallel}^{\sigma\sigma^\prime} + \mathbb{P}_{\parallel}^{\beta\beta^\prime}\mathbb{P}_{\perp}^{\sigma\sigma^\prime}-
\mathbb{P}_{\perp}^{\beta\sigma^\prime}\mathbb{P}_{\parallel}^{\beta^\prime\sigma}-
\mathbb{P}_{\parallel}^{\beta\sigma^\prime}\mathbb{P}_{\perp}^{\beta^\prime\sigma}
\Big),\\
& \mathbb{P}_{1,L}^{\beta\beta^\prime \sigma\sigma^\prime} \equiv \sum_{|J_z|=0}\mathcal {E}_{1,J_z}^{\beta\sigma} \mathcal {E}_{1,J_z}^{\ast \beta ^\prime\sigma^\prime} \nonumber\\ &  = \frac{1}{2}
\Big(  \mathbb{P}_{\perp}^{\beta\beta^\prime}\mathbb{P}_{\perp}^{\sigma\sigma^\prime} -
\mathbb{P}_{\perp}^{\beta\sigma^\prime}\mathbb{P}_{\perp}^{\beta^\prime\sigma}
\Big),
\\
& \mathbb{P}_{2,TT}^{\beta\beta^\prime \sigma\sigma^\prime} \equiv \sum_{|J_z|=2}\mathcal {E}_{2,J_z}^{\beta\sigma} \mathcal {E}_{2,J_z}^{\ast \beta ^\prime\sigma^\prime} \nonumber\\ &  = \frac{1}{2}
\Big(  \mathbb{P}_{\perp}^{\beta\beta^\prime}\mathbb{P}_{\perp}^{\sigma\sigma^\prime} +
\mathbb{P}_{\perp}^{\beta\sigma^\prime}\mathbb{P}_{\perp}^{\beta^\prime\sigma}
\Big)
 - \frac{1}{d-2} \mathbb{P}_{\perp}^{\beta\sigma} \mathbb{P}_{\perp}^{\beta^\prime\sigma^\prime},\\
& \mathbb{P}_{2,T}^{\beta\beta^\prime \sigma\sigma^\prime} \equiv \sum_{|J_z|=1} \mathcal {E}_{2,J_z}^{\beta\sigma} \mathcal {E}_{2,J_z}^{\ast \beta ^\prime\sigma^\prime} \nonumber\\ &  = \frac{1}{2}
\Big(  \mathbb{P}_{\perp}^{\beta\beta^\prime}\mathbb{P}_{\parallel}^{\sigma\sigma^\prime} + \mathbb{P}_{\parallel}^{\beta\beta^\prime}\mathbb{P}_{\perp}^{\sigma\sigma^\prime}+
\mathbb{P}_{\perp}^{\beta\sigma^\prime}\mathbb{P}_{\parallel}^{\beta^\prime\sigma}+
\mathbb{P}_{\parallel}^{\beta\sigma^\prime}\mathbb{P}_{\perp}^{\beta^\prime\sigma}
\Big),\\
& \mathbb{P}_{2,L}^{\beta\beta^\prime \sigma\sigma^\prime} \equiv \sum_{|J_z|=0} \mathcal {E}_{2,J_z}^{\beta\sigma} \mathcal {E}_{2,J_z}^{\ast \beta ^\prime\sigma^\prime} \nonumber\\ &  = \frac{d-2}{d-1}
\Big(
\mathbb{P}_{\parallel}^{\beta\sigma} - \frac{1}{d-2} \mathbb{P}_{\perp}^{\beta\sigma}
\Big)
\Big(
\mathbb{P}_{\parallel}^{\beta^\prime\sigma^\prime} - \frac{1}{d-2} \mathbb{P}_{\perp}^{\beta^\prime\sigma^\prime}
\Big),\\
& \mathbb{P}_{T}^{\beta\beta^\prime \sigma\sigma^\prime} \equiv \sum_{J,J_z,|S_z^\prime|=1} \mathcal {E}_{J,J_z}^{\beta\sigma} \mathcal {E}_{J,J_z}^{\ast \beta ^\prime\sigma^\prime} \nonumber\\ &  = \mathbb{P}^{\beta\beta^\prime} \mathbb{P}_{\perp}^{\sigma\sigma^\prime},\\
& \mathbb{P}_{L}^{\beta\beta^\prime \sigma\sigma^\prime} \equiv \sum_{J,J_z,|S_z^\prime|=0} \mathcal {E}_{J,J_z}^{\beta\sigma} \mathcal {E}_{J,J_z}^{\ast \beta ^\prime\sigma^\prime} \nonumber\\ &  = \mathbb{P}^{\beta\beta^\prime} \mathbb{P}_{\parallel}^{\sigma\sigma^\prime}.
 \end{align}
 \end{subequations}
Where
 \begin{subequations}
    \begin{align}
    \mathbb{P}^{\alpha \alpha '}_\perp  &\equiv \sum_{|S_z| = 1} \epsilon ^\alpha _{S_z}\epsilon ^{*\alpha '}_{S_z}=\sum_{|J_z| = 1} \epsilon ^\alpha _{J_z}\epsilon ^{*\alpha '}_{J_z}=
    -g^{\alpha \alpha '}+\frac{p^\alpha l^{\alpha '}+p^{\alpha '}l^\alpha }{p.l}-\frac{p^2}{(p.l)^2}l^\alpha l^{\alpha '},\\
    \mathbb{P}^{\alpha \alpha '}_\parallel   &\equiv \sum_{|S_z| = 0} \epsilon ^\alpha _{S_z}\epsilon ^{*\alpha '}_{S_z}=\sum_{|J_z| = 0} \epsilon ^\alpha _{J_z}\epsilon ^{*\alpha '}_{J_z}=
    \frac{p^\alpha p^{\alpha '}}{p^2}-\frac{p^\alpha l^{\alpha '}+p^{\alpha '}l^\alpha }{p.l}+\frac{p^2}{(p.l)^2}l^\alpha l^{\alpha '},\\
    \mathbb{P} ^{\alpha \alpha '}& = \sum_{S_z} \epsilon ^\alpha _{S_z}\epsilon ^{*\alpha '}_{S_z}=\sum_{J_z} \epsilon ^\alpha _{J_z}\epsilon ^{*\alpha '}_{J_z}=-g^{\alpha \alpha '}+\frac{p^\alpha p^{\alpha '}}{p^2}.
\end{align}
\end{subequations}

Based on Eq. \eqref{eq:matching}, we can expand $m_Q^2$ in the amplitudes around $M_H^2/4$ before doing the integration for solid angle and the phase space integration when calculating $D_{f\to Q\bar Q[n]}^{NLO}$ and $F^{NLO}_{[n^\prime]\to Q\bar Q[n]}$. Then the calculation is quite similar to that in NRQCD factorization. In Refs. \cite{Chen:2021hzo,Chen:2023gsu} the short-distance hard parts up to $O(\alpha_s^2)$ for $g \to Q\bar Q[\state{{3}}{S}{1}{8}]$, $g \to Q\bar Q[\COaSz]$, $g \to Q\bar Q[\state{{3}}{P}{J,\lambda}{1,8}]$ have been calculated. Following their calculation details, we compute the short distance hard parts for all the single parton FFs, including the gluon FFs, the same quark FFs and different quark FFs. The obtained results are given in the following.

\subsection{Gluon fragmentation functions}

At $O(\alpha_s)$, we have
\begin{subequations}\label{eq:gFFHPLO}
\begin{align}
 \hat{D}_{g \to Q\bar Q[\state{{3}}{S}{1,T}{8}]}^{LO,(0)}( z,M_H, \mu_0, \mu_\Lambda)
 =&  \frac{\pi \alpha_s}{(N_c^2-1)}\frac{8}{M_H^3}  \delta(1-z),
\\
 \hat{D}_{g \to Q\bar Q[\state{{3}}{S}{1,L}{8}]}^{LO,(0)}( z,M_H, \mu_0, \mu_\Lambda)
 =&  0.
\end{align}
\end{subequations}
While all other channels vanish. At $O(\alpha_s^2)$, we have
\begin{subequations}\label{eq:gFFHP}
\begin{align}
& \hat{D}_{g \to Q\bar Q[\state{{3}}{S}{1,T}{1}]}^{NLO,(0)}(z,M_H,\mu_0, \mu_\Lambda)
\nonumber\\=&0,\\
& \hat{D}_{g \to Q\bar Q[\state{{3}}{S}{1,L}{1}]}^{NLO,(0)}(z,M_H,\mu_0, \mu_\Lambda)
\nonumber\\=&0,\\
& \hat{D}_{g \to Q\bar Q[\state{{3}}{S}{1,T}{8}]}^{NLO,(0)}(z,M_H,\mu_0, \mu_\Lambda)
\nonumber\\=&  \frac{4\alpha_s^2N_c }{(N_c^2-1)M_H^3}  \Big[ \frac{1}{2}\delta(1- z)  \Big(2A(\mu_0,M_H)  + \frac{2\beta_0}{N_c}\ln\Big(\frac{\mu_\Lambda^2 e^{ - 1}}{M_H^2}\Big) + \ln^2\Big(\frac{\mu_\Lambda^2 e^{ - 1}}{M_H^2}\Big)
\nonumber\\
&+\frac{\pi^2}{6}
-1 \Big)
+  \frac{1}{N_c} P^{(0)}_{gg}(z) \ln \Big( \frac {\mu_0^2}{\mu_\Lambda^2} \Big)
+  \Big( \frac{2(1-z)}{z} + z( 4+2 z^2)+ \frac{2 z^4}{9}(5+ z) \Big)
\nonumber\\
& \times \Big( \ln\Big( \frac{\mu_\Lambda^2 e^{-1}}{M_H^2}\Big)- 2\ln(1-z) \Big)
 -  \Big( \frac{4z^4}{1-z}- \frac{4z^4}{9}(5+z)\Big) \ln z
 \Big],\\
& \hat{D}_{g \to Q\bar Q[\state{{3}}{S}{1,L}{8}]}^{NLO,(0)}(z,M_H,\mu_0, \mu_\Lambda)
\nonumber\\=& \frac{8\alpha_s^2N_c }{(N_c^2-1)M_H^3} \frac{1-z}{z},\\
& \hat{D}_{g \to Q\bar Q[\state{{1}}{S}{0}{1}]}^{NLO,(0)}( z,M_H, \mu_0, \mu_\Lambda) \nonumber\\ =& \frac{8\alpha_s^2}{M_H^3 N_c}
 \Big[(1-z)\ln(1-z) -z^2 + \frac{3}{2}z \Big],\\
& \hat{D}_{g \to Q\bar Q[\state{{1}}{S}{0}{8}]}^{NLO,(0)}( z,M_H, \mu_0, \mu_\Lambda) \nonumber\\ =& \frac{B_F}{C_F} \hat{D}_{g \to Q\bar Q[\state{{1}}{S}{0}{1}]}^{NLO,(0)}( z,M_H, \mu_0, \mu_\Lambda),\\
& \hat{D}_{g \to Q\bar Q[\state{{1}}{P}{1,T}{1}]}^{NLO,(0)}( z,M_H, \mu_0, \mu_\Lambda) \nonumber\\ =& 0,\\
   & \hat{D}_{g \to Q\bar Q[\state{{1}}{P}{1,L}{1}]}^{NLO,(0)}( z,M_H, \mu_0, \mu_\Lambda) \nonumber\\ =& 0,\\
  & \hat{D}_{g \to Q\bar Q[\state{{1}}{P}{1,T}{8}]}^{NLO,(0)}( z,M_H, \mu_0, \mu_\Lambda) \nonumber\\ =& \frac{32\alpha_s^2}{3M_H^5} \frac{N_c}{(N_c^2-1)z^2}
 \Big[(1-z) \Big(z^3+3z^2-12z+3(3z-4)\ln(1-z)\Big) \Big],\\
   & \hat{D}_{g \to Q\bar Q[\state{{1}}{P}{1,L}{8}]}^{NLO,(0)}( z,M_H, \mu_0, \mu_\Lambda) \nonumber\\ =& \frac{8\alpha_s^2}{3M_H^5} \frac{N_c}{(N_c^2-1)z^2}
 \Big[-2z^4+17z^3-60z^2+48z-6(z^3-7z^2+14z-8)\ln(1-z) \Big],\\
& \hat D^{NLO,(0)}_{g \to Q\bar{Q}[\CScPz]}(z;M_H,\mu_0,\mu_\Lambda)
\nonumber\\ =&
        \frac{32\alpha_s^2}{M_H^5 N_c} \frac{2}{9} \Big[  \frac{1}{36} z (837 - 162 z + 72 z^2 + 40 z^3 + 8 z^4)
 + \frac{9  }{2 } (5 - 3 z ) \ln (1-z) \Big] , \\
& \hat D^{NLO,(0)}_{g \to Q\bar{Q}[\state{{3}}{P}{1,T}{1}]}(z;M_H,\mu_0,\mu_\Lambda) \nonumber\\ =&
    \frac{32 \alpha_s^2}{M_H^5 N_c} \frac{2}{27}   z (9 + 9 z^2 + 5 z^3 + z^4) , \\
& \hat D^{NLO,(0)}_{g \to Q\bar{Q}[\state{{3}}{P}{1,L}{1}]}(z;M_H,\mu_0,\mu_\Lambda) \nonumber\\ =&
\frac{32\alpha_s^2}{M_H^5 N_c} \frac{1}{27} z (9 + 18 z^2 + 10 z^3 + 2 z^4) , \\
& \hat D^{NLO,(0)}_{g \to Q\bar{Q}[\state{{3}}{P}{2,TT}{1}]}(z;M_H,\mu_0,\mu_\Lambda) \nonumber\\ =&
        \frac{32\alpha_s^2}{M_H^5 N_c} \frac{2}{3z^4} \Big[ \frac{2}{9}z (108 - 216 z + 333 z^2 - 225 z^3 + 72 z^4
  + 9 z^6 + 5 z^7
         + z^8)
\nonumber\\
        & - 6(z^5-6 z^4+14 z^3-16 z^2
          +10 z-4 ) \ln(1-z)   \Big] , \\
& \hat D^{NLO,(0)}_{g \to Q\bar{Q}[\state{{3}}{P}{2,T}{1}]}(z;M_H,\mu_0,\mu_\Lambda)
\nonumber\\ =&
        \frac{32\alpha_s^2}{M_H^5 N_c} \frac{1}{3z^4} \Big[  \frac{2}{9}z (-864 + 1728 z - 1368 z^2 + 504 z^3
          - 27 z^4  + 9 z^6 + 5 z^7 + z^8)
\nonumber\\
        &  - 48 (z^4-5 z^3+10 z^2
         -10z+4) \ln(1-z)  \Big] , \\
&\hat D^{NLO,(0)}_{g \to Q\bar{Q}[\state{{3}}{P}{2,L}{1}]}(z;M_H,\mu_0,\mu_\Lambda)
\nonumber\\ =&
        \frac{32\alpha_s^2}{M_H^5 N_c} \frac{1}{9z^4} \Big[  \frac{1}{9}z (3888 - 7776 z + 4212 z^2 - 324 z^3
         - 27 z^4  + 18 z^6 + 10 z^7 +
        2 z^8)
\nonumber\\
        &   - 216(z - 2)(z - 1)^2
          \ln(1-z)  \Big] ,\\
 & \hat D^{NLO,(0)}_{g \to Q\bar{Q}[\state{{3,T}}{P}{}{1}]}(z;M_H,\mu_0,\mu_\Lambda)
\nonumber\\ =&  \frac{32\alpha_s^2}{M_H^5 N_c} \frac{1}{9 z^2}\Big[z (4 z^6+20 z^5+36 z^4+135 z^2
   -126 z+108 )
        -18  (3 z^3-10 z^2+10 z-6 )
\nonumber\\
        &\times \ln (1-z)  \Big], \\
&\hat D^{NLO,(0)}_{g \to Q\bar{Q}[\state{{3,L}}{P}{}{1}]}(z;M_H,\mu_0,\mu_\Lambda)
\nonumber\\ =&  \frac{32\alpha_s^2}{M_H^5 N_c} \frac{1}{2 z^2}\Big[- z(2 z^3 + z^2 - 28 z + 24)
 -2(z - 1) (z^2 + 8 z - 12) \ln (1-z)  \Big], \\
&\hat D^{NLO,(0)}_{g \to Q\bar{Q}[\state{{3}}{P}{J,\lambda}{8}]}(z;M_H,\mu_0,\mu_\Lambda) \nonumber\\ =& \frac{B_F}{C_F} \hat D^{LO,(0)}_{g \to Q\bar{Q}[\state{{3}}{P}{J,\lambda}{1}]}(z;M_H,\mu_0,\mu_\Lambda),\\
&\hat D^{NLO,(0)}_{g \to Q\bar{Q}[\state{{3,\lambda}}{P}{}{8}]}(z;M_H,\mu_0,\mu_\Lambda) \nonumber\\ =& \frac{B_F}{C_F} \hat D^{LO,(0)}_{g \to Q\bar{Q}[\state{{3,\lambda}}{P}{}{1}]}(z;M_H,\mu_0,\mu_\Lambda).
\end{align}
\end{subequations}
Where
\begin{subequations}
\begin{align}
B_F&=\frac{N_c^2-4}{4N_c}, \\
C_F&=\frac{N_c^2-1}{2N_c},\\
A(\mu_0,M_H) &= \frac{\beta_0}{N_c} \Big[ \ln\Big( \frac{\mu_0^2}{M_H^2}\Big) + \frac{13}{3}\Big] + \frac{4}{N_c^2} - \frac{\pi^2}{3} + \frac{16}{3}\ln 2,\\
P^{(0)}_{gg}(z)& =2N_c\Big[ \frac{z}{(1-z)_+} + \frac{1-z}{z} +z(1-z) + \frac{\beta_0}{2N_c} \delta(1-z)\Big],\\
\beta_0 &= \frac{11N_c-2n_f}{6},
\end{align}
\end{subequations}
with $n_f$ denotes the number of light flavors.

\subsection{Same quark fragmentation functions}

For same quark FFs, all channels vanish at $O(\alpha_s)$. At $O(\alpha_s^2)$, we have
\begin{subequations}\label{eq:QFFHP}
\begin{align}
&\hat D^{NLO,(0)}_{Q \to Q\bar{Q}[\state{{3}}{S}{1,T}{1}]}(z;M_H,\mu_0,\mu_\Lambda)
\nonumber\\ =& \frac{64\alpha^2_s C_F^2}{3 M_H^3 N_c}\frac{(z-1)^2}{(z-2)^6}
        z(3z^4-18z^3+38z^2-16z+8),\\
&\hat D^{NLO,(0)}_{Q \to Q\bar{Q}[\state{{3}}{S}{1,L}{1}]}(z;M_H,\mu_0,\mu_\Lambda)
\nonumber\\ =& \frac{16\alpha^2_s C_F^2}{3 M_H^3 N_c}\frac{(z-1)^2}{(z-2)^6}
        z(3z^4-24z^3+64z^2-32z+16),\\
&\hat D^{NLO,(0)}_{Q \to Q\bar{Q}[\state{{3}}{S}{1,T}{8}]}(z;M_H,\mu_0,\mu_\Lambda)
\nonumber\\ =&
\frac{2\alpha^2_s}{N_cM_H^3z} \Big[ \frac{-z^4+10z^3-18z^2+16z-8}{(z-2)^2}+ (z^2-2z+2)\ln
\frac{4\mu_0^2}{(2-z)^2 M_H^2}
\nonumber\\
&
+\frac{8}{3N_c^2(z-2)^6}
z(1-z)
 \Big(3N_c(z^3-7z^2+8z-4)(z-2)^2+z(-3z^5+21z^4-56z^3
\nonumber\\
&
+54z^2-24z+8)\Big) \Big],\\
&\hat D^{NLO,(0)}_{Q \to Q\bar{Q}[\state{{3}}{S}{1,L}{8}]}(z;M_H,\mu_0,\mu_\Lambda)
\nonumber\\ =& \frac{2\alpha^2_s}{N_c M_H^3z}
        \Big[\frac{8(z-1)^2}{(z-2)^2}-\frac{2}{3N_c^2(z-2)^6}(1-z)^2z^2 \Big(12N_c(z-4)
        (z-2)^2-3z^4+24z^3
        \nonumber\\
        & -64z^2+32z-16 \Big) \Big],\\
&\hat D^{NLO,(0)}_{Q \to Q\bar{Q}[\state{{1}}{S}{0}{1}]}(z;M_H,\mu_0,\mu_\Lambda)
\nonumber\\ =& \frac{16}{3 M_H^3}\frac{C^2_F\alpha^2_s}{N_c}\frac{(z-1)^2}{(z-6)^2}
        z(3z^4-8z^3+8z^2+48),\\
&\hat D^{NLO,(0)}_{Q \to Q\bar{Q}[\state{{1}}{S}{0}{8}]}(z;M_H,\mu_0,\mu_\Lambda)
\nonumber\\ =& \frac{1}{(N_c^2-1)^2}\hat D^{NLO,(0)}_{Q \to Q\bar{Q}[\state{{1}}{S}{0}{1}]}(z;M_H,\mu_0,\mu_\Lambda),\\
&\hat D^{NLO,(0)}_{Q \to Q\bar{Q}[\state{{1}}{P}{1,T}{1}]}(z;M_H,\mu_0,\mu_\Lambda)
\nonumber\\ =& \frac{512}{15 M_H^5}\frac{C^2_F\alpha^2_s}{N_c}\frac{(z-1)^2}{(z-2)^8}
z(10z^6-76z^5+233z^4-328z^3+256z^2-160z+80), \\
&\hat D^{NLO,(0)}_{Q \to Q\bar{Q}[\state{{1}}{P}{1,L}{1}]}(z;M_H,\mu_0,\mu_\Lambda)
\nonumber\\ =&  \frac{64}{15 M_H^5}\frac{C^2_F\alpha^2_s}{N_c}\frac{(z-1)^2}{(z-2)^8}
z(55z^6-232z^5+236z^4+224z^3+592z^2-640z+320),\\
&\hat D^{NLO,(0)}_{Q \to Q\bar{Q}[\state{{3}}{P}{0}{1}]}(z;M_H,\mu_0,\mu_\Lambda)
\nonumber\\ =&
\frac{64}{9 M_H^5}\frac{C^2_F\alpha^2_s}{N_c}\frac{(z-1)^2}{(z-2)^8}
        z(59z^6-376z^5+1060z^4-1376z^3+528z^2+384z+192),\\
&\hat D^{NLO,(0)}_{Q \to Q\bar{Q}[\state{{3}}{P}{1,T}{1}]}(z;M_H,\mu_0,\mu_\Lambda)
\nonumber\\ =& \frac{128}{15M_H^5}\frac{C^2_F\alpha^2_s}{N_c}\frac{(z-1)^2}{(z-2)^8}
        z(35z^6-228z^5+884z^4-2064z^3+3088z^2-1920z+640),\\
&\hat D^{NLO,(0)}_{Q \to Q\bar{Q}[\state{{3}}{P}{1,L}{1}]}(z;M_H,\mu_0,\mu_\Lambda)
\nonumber\\ =&
\frac{128}{15 M_H^5}\frac{C^2_F\alpha^2_s}{N_c}\frac{(z-1)^2}{(z-2)^8}
        z(35z^6-312z^5+1136z^4-2016z^3+1872z^2-960z+320),\\
&\hat D^{NLO,(0)}_{Q \to Q\bar{Q}[\state{{3}}{P}{2,TT}{1}]}(z;M_H,\mu_0,\mu_\Lambda)
\nonumber\\ =&
\frac{32}{15 M_H^5}\frac{32C^2_F\alpha^2_s}{N_c}\frac{(z-1)^4}{(z-2)^8}
        z(5z^4-32z^3+68z^2-32z+16),\\
&\hat D^{NLO,(0)}_{Q \to Q\bar{Q}[\state{{3}}{P}{2,T}{1}]}(z;M_H,\mu_0,\mu_\Lambda)
\nonumber\\ =&
\frac{32}{15 M_H^5}\frac{4C^2_F\alpha^2_s}{N_c}\frac{(z-1)^2}{(z-2)^8}
        z(75z^6-580z^5+1628z^4-1872z^3+1328z^2-512z+128),\\
&\hat D^{NLO,(0)}_{Q \to Q\bar{Q}[\state{{3}}{P}{2,L}{1}]}(z;M_H,\mu_0,\mu_\Lambda)
\nonumber\\ =&
\frac{32}{45M_H^5}\frac{4C^2_F\alpha^2_s}{N_c}\frac{(z-1)^2}{(z-2)^8}
        z(115z^6-932z^5+2648z^4-2944z^3+2064z^2-768z+192),\\
&\hat D^{NLO,(0)}_{Q \to Q\bar{Q}[\state{{3,T}}{P}{}{1}]}(z;M_H,\mu_0,\mu_\Lambda)
\nonumber\\ =&
\frac{128}{3M_H^5}\frac{C^2_F\alpha^2_s}{N_c}\frac{(z-1)^2}{(z-2)^8}z(43z^6-320z^5
+964z^4-1376z^3+1168z^2-512z+192),\\
&\hat D^{NLO,(0)}_{Q \to Q\bar{Q}[\state{{3,L}}{P}{}{1}]}(z;M_H,\mu_0,\mu_\Lambda)
\nonumber\\ =&
\frac{64}{3M_H^5}\frac{C^2_F\alpha^2_s}{N_c}\frac{(z-1)^2}{(z-2)^8}z(23z^6-192z^5
+676z^4-1120z^3+1104z^2-512z+192),\\
&\hat D^{NLO,(0)}_{Q \to Q\bar{Q}[\state{{1}}{P}{1,\lambda}{8}]}(z;M_H,\mu_0,\mu_\Lambda)
\nonumber\\ =&
\frac{1}{(N_c^2-1)^2} \hat D^{NLO,(0)}_{Q \to Q\bar{Q}[\state{{1}}{P}{1,\lambda}{1}]}(z;M_H,\mu_0,\mu_\Lambda),
\\
&\hat D^{NLO,(0)}_{Q \to Q\bar{Q}[\state{{3}}{P}{J,\lambda}{8}]}(z;M_H,\mu_0,\mu_\Lambda)
\nonumber\\ =&
\frac{1}{(N_c^2-1)^2} \hat D^{NLO,(0)}_{Q \to Q\bar{Q}[\state{{3}}{P}{J,\lambda}{1}]}(z;M_H,\mu_0,\mu_\Lambda),
\\
&\hat D^{NLO,(0)}_{Q \to Q\bar{Q}[\state{{3,\lambda}}{P}{}{8}]}(z;M_H,\mu_0,\mu_\Lambda)
\nonumber\\ =&
\frac{1}{(N_c^2-1)^2}\hat D^{NLO,(0)}_{Q \to Q\bar{Q}[\state{{3,\lambda}}{P}{}{1}]}(z;M_H,\mu_0,\mu_\Lambda).
\end{align}
\end{subequations}

\subsection{Different quark fragmentation functions}

The short distance hard parts for different quark FFs receive contributions that
begin at $O(\alpha_s^2)$, which read
\begin{subequations}\label{eq:qFFHP}
\begin{align}
&\hat D^{NLO,(0)}_{Q^\prime \to Q\bar{Q}[\state{{3}}{S}{1,T}{8}]}(z;M_H,\mu_0,\mu_\Lambda)
\nonumber\\ =&
\frac{2\alpha^2_s}{N_c M_H^3z}
\Big[ -\frac{(z^4-2z^3+2z^2)\eta+8z^3-16z^2+16z-8}{\eta z^2-4z+4}
 + (z^2-2z+2)
 \nonumber\\
& \times \ln \Big(\frac{\mu^2_0}{M_H^2(1-z+z^2\eta/4)}\Big) \Big],\\
&\hat D^{NLO,(0)}_{Q^\prime \to Q\bar{Q}[\state{{3}}{S}{1,L}{8}]}(z;M_H,\mu_0,\mu_\Lambda)
\nonumber\\ =& \frac{16\alpha^2_s}{N_c M_H^3z}\frac{(z-1)^2}{\eta z^2-4z+4},\\
&\hat D^{NLO,(0)}_{Q^\prime \to Q\bar{Q}[n]}(z;M_H,\mu_0,\mu_\Lambda)
\nonumber\\ =& 0 \qquad (n\neq \state{{3}}{S}{1,\lambda}{8}).
\end{align}
\end{subequations}
Here quark $Q^\prime$ has a different flavor with outgoing $Q\bar Q$ pair.
And $\eta=4m_{Q^\prime}^2/M_H^2$ with $m_{Q^\prime}$ denotes the mass of quark $Q^\prime$. When $m_{Q^\prime}$ is the light quark mass, $\eta =0$.

In the calculation, we find our results for the perturbative calculated FFs are agree with that in Refs. \cite{Ma:2013yla,Ma:2014eja,Ma:2015yka,Bodwin:2014bia}. On the other hand, we find that all infrared (IR) divergences are canceled at $O(\alpha_s^2)$, and all derived short distance
hard parts are finite. Besides, we find the hard parts are also free of
the plus distributions with the scale choice $\mu_0=\mu_\Lambda=M_H$.

\section{SUMMARY}\label{sec:summary}

In this paper we studied the single parton FFs of heavy quarkonium in SGF approach. In SGF, the FFs are expressed as the convolution of perturbative short distance
hard parts with the SGD in Eq. \eqref{eq:SGF-form}. We calculated the short distance
hard parts for all single parton FFs up to $O(\alpha_s^2)$ by matching the perturbative calculated FFs to the perturbative calculated SGDs. In our calculations, we obtained FFs that agree with the results in literature. Notably, we have found that all IR divergences cancel at this order, resulting in finite short-distance hard parts. Furthermore, by choosing appropriate natural scales, the derived short-distance hard parts are also free of plus distributions. These results establish the viability of using SGF to study heavy quarkonium production at high-energy colliders and within jets. \cite{Baumgart:2014upa,Kang:2017yde,Bain:2017wvk,LHCb:2017llq}.

\section*{Acknowledgments}

This work is supported by the National Natural Science Foundation of China (Grants No. 12205124 and No. 12365014).


\providecommand{\href}[2]{#2}\begingroup\raggedright\endgroup


\end{document}